\documentclass[conference]{IEEEtran}
\IEEEoverridecommandlockouts
\usepackage{cite}
\usepackage{amsmath,amssymb,amsfonts}
\usepackage{algorithmic}
\usepackage{graphicx}
\usepackage{textcomp}
\usepackage{xcolor}
\def\BibTeX{{\rm B\kern-.05em{\sc i\kern-.025em b}\kern-.08em
    T\kern-.1667em\lower.7ex\hbox{E}\kern-.125emX}}

\usepackage{siunitx}
\newcommand{\lowr}[1] {_{\mathrm{#1}}}

\begin{document}

\title{Towards Real Time Thermal Simulations for Design Optimization using Graph Neural Networks}
\author{
\IEEEauthorblockN{1\textsuperscript{st} H\`elios Sanchis-Alepuz}
\IEEEauthorblockA{\textit{Power Electronics} \\
\textit{Silicon Austria Labs GmbH}\\
Graz, Austria \\
helios.sanchis-alepuz@silicon-austria.com}
\and
\IEEEauthorblockN{2\textsuperscript{nd} Monika Stipsitz}
\IEEEauthorblockA{\textit{Power Electronics} \\
\textit{Silicon Austria Labs GmbH}\\
Graz, Austria \\
monika.stipsitz@silicon-austria.com}
}

\maketitle

\begin{abstract}
This paper presents a method to simulate the thermal behavior of 3D systems using a graph neural network. The method discussed achieves a significant speed-up with respect to a traditional finite-element simulation. The graph neural network is trained on a diverse dataset of 3D CAD designs and the corresponding finite-element simulations, representative of the different geometries, material properties and losses that appear in the design of electronic systems. We present for the transient thermal behavior of a test system. The accuracy of the network result for one-step predictions is remarkable (\SI{0.003}{\%} error). After 400 time steps, the accumulated error reaches \SI{0.78}{\%}. The computing time of each time step is \SI{50}{ms}. Reducing the accumulated error is the current focus of our work. In the future, a tool such as the one we are presenting could provide nearly instantaneous approximations of the thermal behavior of a system that can be used for design optimization.
\end{abstract}

\begin{IEEEkeywords}
3D geometries, Graph Neural Network, Transient heat equation
\end{IEEEkeywords}

\section{Introduction}
In the design of electronic systems, the ``Simulation-Based Design'' (SBD) paradigm aims at using simulations of the behavior of a system as soon as possible in the design process in order to minimize design costs and eliminate unfit designs as early as possible. This includes not only electronic circuit simulations but also simulations of the physical behavior (thermal, mechanical, etc.) of the system. In the field of Power Electronics, a particularly important design objective is the optimization of the thermal performance of the final system. It would be extremely beneficial to have an automatized design process in which many different layouts could be studied and those offering, e.g., optimal thermal behavior, size, cost, etc. be selected.

The main obstacle hindering a more widespread use of SBD is the long simulation time associated with the traditional simulation methods such as Finite Element methods (FEM). Early evaluation via simulations of potential designs requires considerable computational resources and is often unfeasible. This paper presents the implementation of a learned simulator \cite{Grzeszczuk1998,Sanchez-Gonzalez2020} using a graph neural network (GNN) \cite{Scarselli2009,Battaglia2018}. A learned graph-network simulator (GNS) infers, from a representative training dataset, the dynamics of the physical phenomenon of interest (here, heat propagation) and encodes it by means of a neural network. The GNS can then be applied to any other system and provide a simulation result typically much faster than a traditional simulator would.

\begin{figure}[h]
    \centerline{\includegraphics[width=\columnwidth]{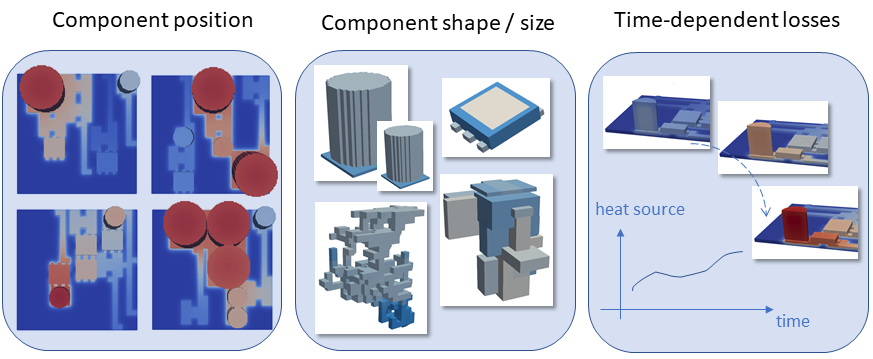}}
    \caption{The design space that the network should learn during training: (1) Components located at varying positions on the PCB, (2) Different component shapes and sizes (including completely random component geometries), (3) Heat sources varying over time.}
    \label{fig:design_space}
\end{figure}

This paper shows results of a GNS for the transient thermal behavior of 3D systems representing electronic systems, as shown in Fig.~\ref{fig:design_space}. The training dataset consists of a large number of randomly generated 3D CAD systems with different material properties assigned to different parts of the system and heat sources representing the losses in an electronic circuit. The result of a thermal FEM simulation on each of those systems is the ground truth from which the GNS must learn the dynamics of heat propagation. As a result of the GNS learning the correct dynamics of heat, it is possible to train the GNS on 3D systems that are much simpler than a realistic electronic system, in contrast to the approach in \cite{Stipsitz2022}; this is one of the main novelties of this work.

\section{Graph Neural Networks}\label{sec:GNNs}

\begin{figure}[h]
    \centerline{\includegraphics[width=\columnwidth]{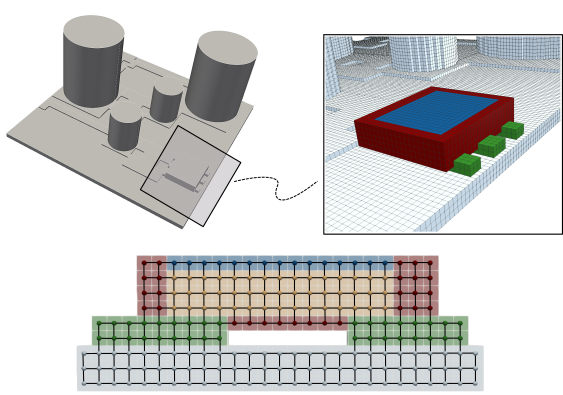}}
    \caption{Illustration of the graph generation. First, the components in the CAD design (left) are split into cubes (right). In the graph each cubic volume is represented by a node. The image in the bottom shows a 2D section of the chip geometry and the corresponding graph in overlay.}
    \label{fig:graph_generation}
\end{figure}

A GNN is designed to map an input graph onto an output graph. Here, a graph is a collection $(N,E)$ of nodes $\mathbf{n_i}\in N$ with certain attributes, connected via edges $\mathbf{e_{ij}}\in E$ which may also have attributes. In order to apply that methodology to our problem, a regular discretization is applied on the 3D systems (see Fig.~\ref{fig:graph_generation}). Each element of the discretized system is represented by a node, whose attributes are related to the heat conductivity, density, and heat capacity of the corresponding material, as well as the initial average temperature of that element; in order to represent losses in electronic circuits, some nodes have as an additional attribute a non-vanishing heat source value (see Eqs.~\eqref{Eq:update_heateq} or \eqref{Eq:update_graph}). The interfaces between the elements are represented as edges. The collection of those nodes and edges constitutes the input graph. After applying the GNS, the output graph has the same node-edge structure(i.e. the same number of nodes and edges, with the same connectivity as the input graph), but with the average temperature of each node updated to its value after a certain time step (in this study, after \SI{0.01}{s}). By recursively applying the GNS, we obtain the full time-dependent thermal evolution of the system.

\begin{figure*}[h]
    \centerline{\includegraphics[width=0.8\textwidth]{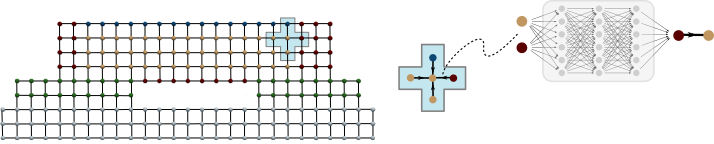}}
    \caption{The temperature at each node is updated by accumulating fluxes sent from the neighboring nodes (indicated by a blue cross in the left image). The fluxes between two neighboring nodes are approximated by a multilayer perceptron (right).}
    \label{fig:graph}
\end{figure*}

\subsection{Structure of the GNS}\label{sec:gns_structure}

In this work we implement a simplified version of the message-passing GNS framework put forward in \cite{Sanchez-Gonzalez2020}, which maps the input graph onto the output graph by successively applying three computational blocks: encoder, processor and decoder. In our simplified version, the GNS is composed of encoder and decoder blocks only.

{\bf Encoder}. The encoder maps the input graph onto a graph of identical node-edge structure called latent graph, but where the latent nodes and edges have a different (larger) number of attributes than the input graph. The latent edge attributes $\mathbf{e^{\textnormal{lat.}}_{ij}}$ are obtained via the application of an edge embedding function $\mathbf{e^{\textnormal{lat.}}_{ij}}=f^e_{\textnormal{lat.}}\left(\mathbf{n_i},\mathbf{n_j}\right)$ on all pairwise connected nodes $\mathbf{n_i}$, $\mathbf{n_j}$ of the input graph. The attributes of the latent nodes $\mathbf{n^{\textnormal{lat.}}_{i}}$  are similarly obtained via the application of a node embedding function $\mathbf{n^{\textnormal{lat.}}_{i}}=f^n_{\textnormal{lat.}}\left(\mathbf{n_i},aggr(e^{\textnormal{lat.}})\right)$, where $aggr(e)$ is a fixed edge aggregation function, which in our case simply adds up element-wise all the edges attached to each node $n^{\textnormal{lat.}}_{i}$. The embedding functions $f^e_{\textnormal{lat.}}$ and $f^n_{\textnormal{lat.}}$ are learned functions, parametrized by multilayer perceptrons (MLPs).

{\bf Decoder}. The decoder then maps the latent graph onto the output graph as follows. In the output graph, the edges have a single attribute $F_{ij}$ which is expected to represent the heat flux from node $j$ into node $i$, as illustrated in Fig.~\ref{fig:graph}. The flux is obtained by the application of an edge function $\mathbf{F_{ij}}=f^e\left(n^{\textnormal{lat.}}_{i},n^{\textnormal{lat.}}_{i}\right)$ on the latent graph. As in the encoder block, the function $f^e$ is a learned function parametrized by an MLP. The node attributes of the output graph are the material properties (and possible heat source values) of the element represented by the node as well as the updated average temperature, given by\footnote{This update function is obtained from an integral form of the heat equation
\begin{equation}
    \left<T_i\right>\lowr{t+\Delta t} = \left<T_i\right>\lowr{t} + \frac{\Delta t}{V_i \rho_i c_{\mathrm{P},i}}\int_\Sigma k_i\nabla T d\Sigma + \frac{h_i}{c_{\mathrm{P},i}}~,
\label{Eq:update_heateq}
\end{equation}
with $k_i$ the heat conductivity, under the assumption that the edge attribute $F_{ij}$ faithfully approximates the surface integral of the heat flux $k\nabla T$ across the interface between two discretization elements, such that we can replace the surface integral $\int_\Sigma k_i\nabla T d\Sigma$ by the sum $\sum_j F_{ij}$. In both \eqref{Eq:update_heateq} and \eqref{Eq:update_graph}, the time derivative of the temperature field has been replaced by a finite-difference approximation.}
\begin{equation}
    \left<T_i\right>\lowr{t+\Delta t} = \left<T_i\right>\lowr{t} + \frac{\Delta t}{V \rho_i c_{\mathrm{P},i}}\sum_j F_{ij} + \frac{h_i}{c_{\mathrm{P},i}}~,
\label{Eq:update_graph}
\end{equation}
where $\left<T_i\right>\lowr{t}$ is the volume-average temperature in node $i$ at time $t$, $\rho_i$ and $c_{\mathrm{P},i}$ are, respectively, the density and heat capacity of the material represented by node $i$ and $V$ is the volume of the discretization element. As we said above, possible electronic losses are represented by heat sources of value $h_i$. Note that the material properties as well as $h_i$ are assumed to be constant within a given discretization element (viz. for a given node) and time step but they are permitted to change from element to element and from one time step to the next.

{\bf Boundary conditions}. We consider two possible boundary conditions (BCs) for the heat equation, namely Dirichlet conditions where the temperature is fixed at certain surfaces and Neumann conditions where the heat flux across certain surfaces is determined by
\begin{equation}
k~\nabla T\cdot \hat{n}=\alpha~(T_{BC}-T)~,
 \label{eq:NeumannBC}
\end{equation}
with $T_{BC}$ some fixed temperature, $\alpha$ a heat transfer coefficient and $\hat{n}$ the outwards normal unit vector of the corresponding surface. In our GNS, these BCs are implemented by introducing special auxiliary nodes (see Sec.~\ref{sec:implementation}). For these nodes, there is no temperature update.

\subsection{Implementation details}\label{sec:implementation}

The attributes of the nodes in the input graph are
\begin{equation}
\textnormal{node attributes}=\left\{\frac{k}{\rho c_P},\frac{1}{\rho c_P},V^{1/3},\frac{\left<T\right>}{1000},\rho h V \right\}~.
\end{equation}
Since the input attributes are all positive but with very broad range of values, we normalize them as $x\rightarrow -\log_{10}(x)/10$.
For the auxiliary boundary condition nodes, the attributes are instead
\begin{equation}
\textnormal{Dirichlet node attributes}=\left\{0,0,V^{1/3},\frac{\left<T_{BC}\right>}{1000},0\right\}~,
\end{equation}
\begin{equation}
\textnormal{Neumann node attributes}=\left\{\alpha,0,V^{1/3},\frac{\left<T_{BC}\right>}{1000},0\right\}~,
\end{equation}
and the normalization is skipped for the vanishing attributes.

The encoder implements the embedding functions $f^e_{\textnormal{lat.}}$ and $f^n_{\textnormal{lat.}}$ as MLPs with two hidden layers, with 128 units each, and SELU activation functions \cite{klambauer2017self}, followed by a non-activated output layer with 128 units. Hence, the nodes and edges of the resulting latent graph have 128 attributes each. We use LayerNorm \cite{ba2016layer} after each hidden layer of the MLPs to improve training stability.

The decoder implements the edge function $f^e$ also as an MLP with two hidden layers, with 128 units each and each followed by a LayerNorm. In this case we use the ReLU activation function. The output layer contains a single unit and is followed by a $\textrm{Sinh}$ function. This is because the numerical value of the heat flux inside a system can vary across several orders of magnitudes. Passing the output of the MLP through a $\textrm{Sinh}$ allows the network to output values within a large range whilst still keeping the weights numerically small.

In our architectural choices we took advantage of the extensive study in \cite{Sanchez-Gonzalez2020}. Starting with the optimal hyperparameters found therein, we performed a modest architecture exploration before settling on the choices stated above.

\subsection{Training details}\label{sec:training}

\begin{figure*}[h]
    \centerline{\includegraphics[width=0.8\textwidth,trim=0 60 200 35,clip]{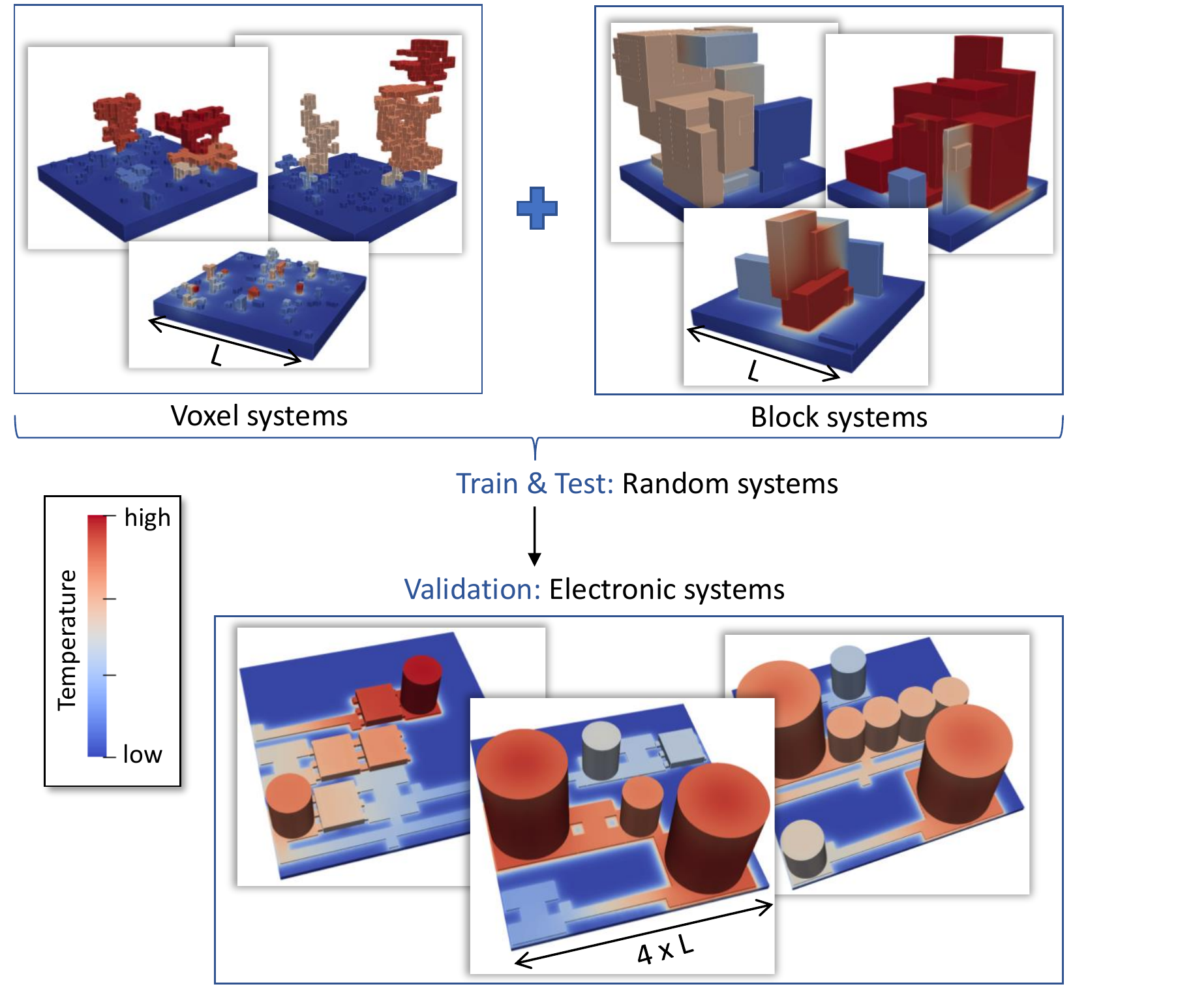}}
    \caption{The different types of 3D systems generated in this work. The random systems (voxel systems and block systems) are used for training and testing. The validation of the model is performed using the electronic systems, which are four times larger per dimension than the random systems.}
    \label{fig:systems}
\end{figure*}

{\bf Dataset}. Even though the envisioned application of our GNS is the fast thermal simulation of electronic systems, for the training of the GNS a set of 3D systems with random geometries are used (see Figs.~\ref{fig:design_space} and \ref{fig:systems}). The training is performed using supervised learning with transient FEM simulations of those systems as ground truth. We have previously observed that training on random systems lead to better generalization capabilities of a learned simulator to previously unseen geometries \cite{StipsitzEccomas2022}. Additionally, these random geometries are much simpler to generate automatically than electronic systems.

Two types of random systems were generated to cover the variety of geometrical shapes found in electronic systems: ``voxel systems'' constructed of individual voxels where each one is assigned a different random material, and ``block systems'' made up of larger boxes leading to larger regions with the same material properties. To increase the variation in geometrical shapes, the block systems were generated from randomly sized 3D blocks which were randomly placed on a surface (akin to a PCB, in order to ensure thermal connectivity between blocks). The blocks could be overlapping, in which case the structure was decomposed into multiple non-overlapping segments. For more details on the different datasets see \cite{StipsitzEccomas2022}.

In total, 200 of these random systems were generated. The first \SI{4}{s} of the time evolution of the temperature of each system was simulated in \SI{0.01}{s} steps. Of these, 375 out of 400 time steps for each system were randomly selected for training. \SI{80}{\%} of the systems and time steps were used for training, the rest for testing.

To validate the generalization capability of our GNS, an additional set of electronic systems was generated. Each system consists of a PCB on which electronic components (IC, large and small capacitors and copper layers of different shapes and sizes) are placed at random locations.

In all cases, the material properties were varied in the following ranges: $k = \SIrange{0.66}{1.1}{W/(Km)}$, $\rho = \SIrange{1261.5}{2102.5}{kg/m^3}$, $c\lowr{P} = \SIrange{714}{1190}{J/(kg K)}$ and heat source powers per voxel of $\rho h V = \SIrange{0}{6e-4}{W}$. Dirichlet BCs were applied on the bottom of the systems with boundary temperatures varying in the range \SIrange{280}{400}{K}. To every other exposed surface we applied Neumann BCs with varying $\alpha$ of \SIrange{10}{20}{W/(Km^2)} and the same BC temperature; this imitates heat rejection into air, whilst keeping heat conduction as the only physical domain under consideration.

From each 3D system, a graph is created as explained in Sec.~\ref{sec:GNNs}. Specifically, the systems are discretized using a regular cubic grid of \SI{0.2}{mm} grid size (i.e. with cells of size \SI{0.2}{mm}$\times$\SI{0.2}{mm}$\times$\SI{0.2}{mm}), see Fig.~\ref{fig:graph_generation}.

All systems were generated in an automatic workflow including CAD generation (with FreeCAD \cite{freecad}), meshing (with gmsh \cite{geuzaine2009gmsh}), FEM simulation (with Elmer \cite{elmer}) and postprocessing to convert the results to a graph structure.

{\bf Loss function and optimization}. The GNS is trained on single-step pairs $(\left<T_t^{\textnormal{FEM}}\right>,\left<T_{t+\Delta t}^{\textnormal{FEM}}\right>)$ and the model is optimized to minimize a relative $L_1$ loss
\begin{equation}
L_1^{rel}=\frac{||\left<T_{t+\Delta t}^{\textnormal{GNS}}\right>-\left<T_{t+\Delta t}^{\textnormal{FEM}}\right>||}{\left<T_{t+\Delta t}^{\textnormal{FEM}}\right>}~.
 \label{eq:L1loss}
\end{equation}
We used the Adam optimizer \cite{kingma2014adam} with a mini-batch size of 10. The learning rate was decreased from $10^{-4}$ to $10^{-6}$ using a fixed-step scheduler. The training typically involved around 5.6M gradient updates. On an Nvidia Titan RTX GPU each training epoch took approximately \SI{14}{min}, the training typically requiring between 700 and 1000 epochs.

During training, we corrupt the input temperatures with multiplicative Gaussian noise $\mathcal{N}(1,0.00003)$. In \cite{Sanchez-Gonzalez2020} this was suggested as an approach to mitigate the effect of error accumulation during thermal trajectory rollouts (see Sec.~\ref{sec:results}). In contrast to their observation, however, we have not noticed any significant improvement on the accumulated rollout error compared to uncorrupted input temperatures. Nevertheless, we still use noise injection during training (not during evaluation), as it seems to prevent the MLPs from overfitting, and deactivate it in the last stages of training.

The GNN models were implemented using PyTorch Geometric \cite{Fey/Lenssen/2019} and the training was implemented using PyTorch \cite{NEURIPS2019_9015}.

\section{Results}\label{sec:results}

The GNS can predict the temperature map at the next time step based on the previous time step. 
For the test set formed by
``voxel systems'' and ``block systems'', the average $L_1^{\textnormal{rel}}$ error for such a one-step prediction is \SI{0.002}{\%}. A much more stringent test of the performance of the GNS is the accumulated error in a rollout trajectory: the system starts in a homogeneous temperature state and over time heats up due to heat sources; the first prediction of the network is based on the initial temperature distribution only; subsequent temperature predictions are based on previous predictions of the GNS so that errors accumulate over time. For rollout trajectories in the test set, the average accumulated error after 400 time steps of \SI{0.01}{s} is \SI{0.25}{\%}.

\begin{figure*}[h]
    \centerline{\includegraphics[width=\textwidth,trim=0 0.5cm 0 2.5cm,clip]{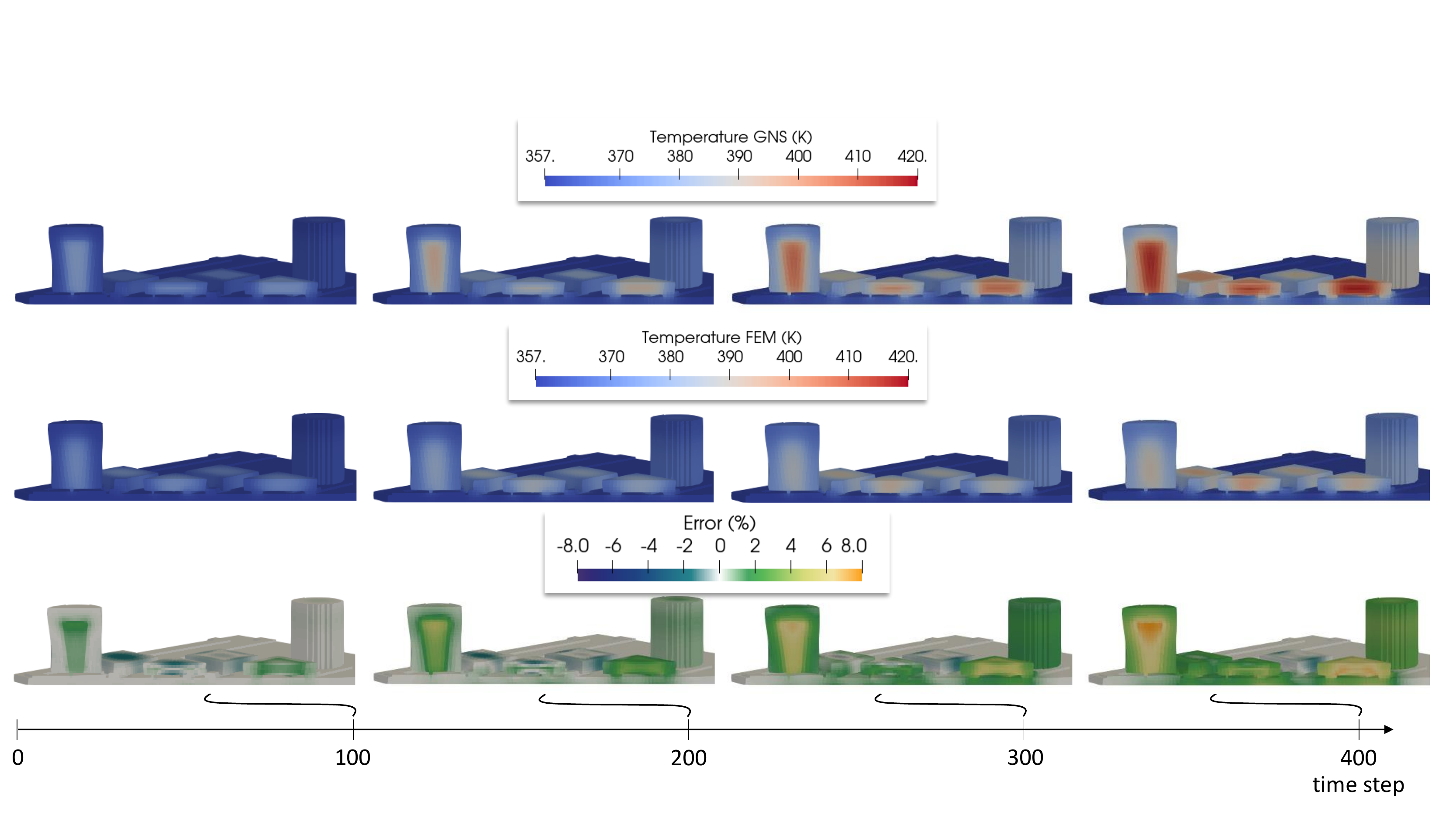}}
    \caption{Temperature distribution visualized on a cross-section of a selected electronic system as obtained from unrolling the GNS (first row) compared to the baseline temperature of the FEM simulation (second row). The system starts at a uniform temperature (at $t=0$) and in the succeeding time steps heats up due to heat sources at the center of the chips and capacitors. The error $\frac{\left<T_{t+\Delta t}^{\textnormal{GNS}}\right>-\left<T_{t+\Delta t}^{\textnormal{FEM}}\right>}{\left<T_{t+\Delta t}^{\textnormal{FEM}}\right>}$ (in percent in third row) is highest at the location of the heat sources.}
    \label{fig:unrolling}
\end{figure*}

\begin{figure}[h]
    \centering
    \begin{minipage}{0.5\columnwidth}
        \includegraphics[width=\textwidth]{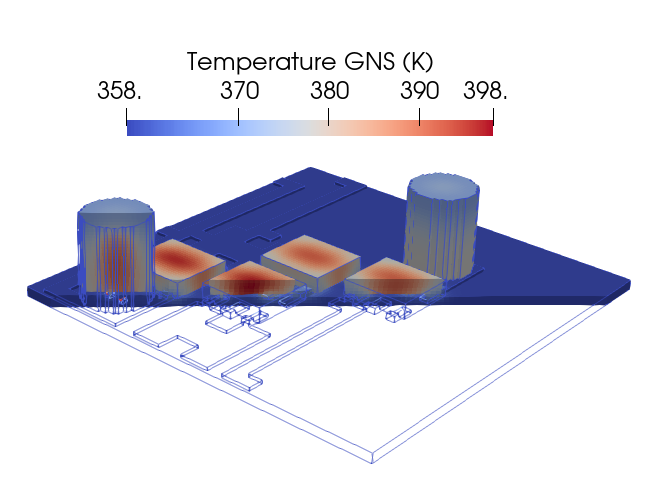}
    \end{minipage}%
    \begin{minipage}{0.5\columnwidth}
        \includegraphics[width=\textwidth]{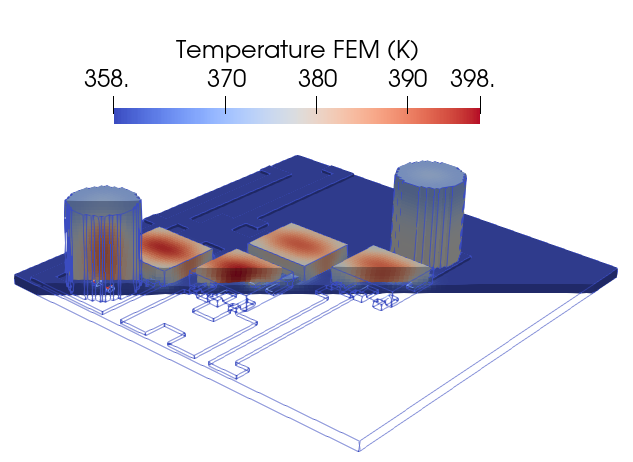}
    \end{minipage}
    \caption{The one-step temperature prediction from the GNN (left) and the ground truth obtained from a FEM simulation (right) are indistinguishable.}
    \label{fig:onestep_result}
\end{figure}

For the generalization analysis of our GNS we evaluate the model on the generated set of electronic systems, unseen during training. These systems are 4 times larger per dimension than the systems in the training dataset (see Fig.~\ref{fig:systems}). For those systems, the average one-step prediction error is \SI{0.003}{\%}, which is still remarkably low (see Fig.~\ref{fig:onestep_result}).

\begin{figure}[h]
    \centerline{\includegraphics[width=0.8\columnwidth]{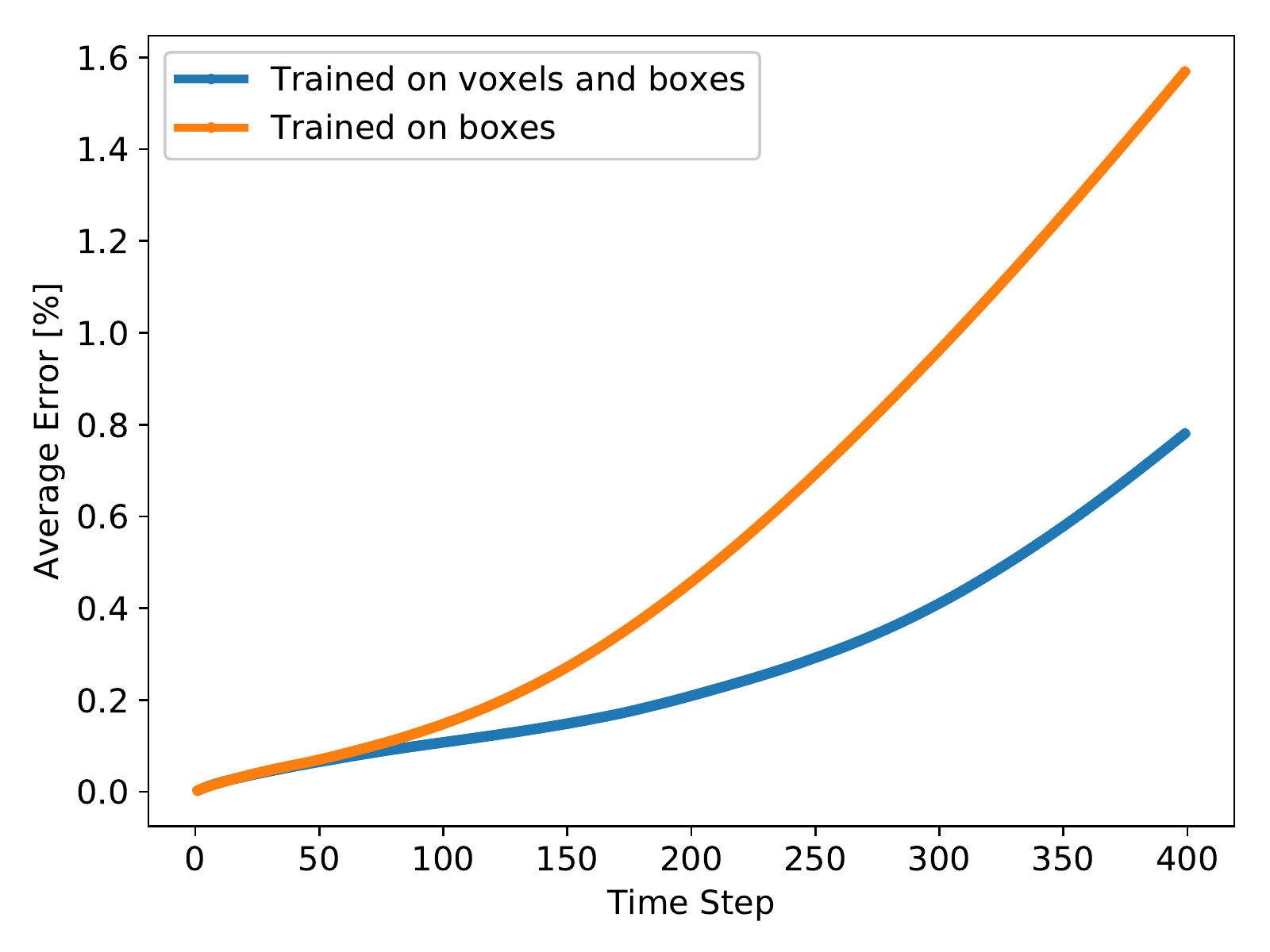}}
    \caption{The average rollout error evaluated for the set of electronic systems depends strongly on the type of systems used for training the GNS. A training dataset combining voxel and box systems (cf.\ Fig.~\ref{fig:systems}) leads to better generalization to the electronic systems compared to training only on box systems.}
    \label{fig:unrolling_error}
\end{figure}

In Fig.~\ref{fig:unrolling_error} we show several snapshots of the rollout trajectory of one of the electronic systems. For this particular system, the accumulated average error after 400 time steps reaches $1.4\%$. Locally, however, the accumulated error can be larger as visible in Fig.~\ref{fig:unrolling_error}. The largest local errors occur at the location of the heat sources (namely, the central region of the components). This is likely due to the fact that, at the interface between the regions with and without heat sources, the heat flux can be very large and the MLP still has problems generating numerically large output values; too small a heat flux in the GNS implies that not enough heat is leaving the regions with heat sources and, hence, they get too hot over time.

Finally, it is interesting to mention that training on the combination of ``voxel'' and ``block systems'' gives significantly better results than training on one of them only, as was already observed in \cite{StipsitzEccomas2022}. We show in Fig.~\ref{fig:unrolling_error} a comparison of the average rollout error for the electronic systems when the GNS is trained on ``block systems'' only, compared to the case where the GNS was trained on both datasets. The reason for the improvement is that the very different geometries generate also a more diverse landscape of heat fluxes, which results in a better training dataset. Obviously, this opens the question of whether even better training systems could be designed; this is the subject of ongoing work.

%
Overall, the GNS achieves a remarkable accuracy keeping in mind that the approach applies a much coarser spatial and time discretization than would be appropriate with conventional methods. With regards to evaluation time of the GNS, for the electronic systems each time step is evaluated in \SI{50}{ms} which is a substantial speed-up compared to the FEM simulations which, on average, needed \SI{300}{s} per time step. Even though the comparison is not entirely fair since for the FEM simulations the open-source tool ElmerFEM was run on a single thread of an Intel Xeon W-2145 CPU, while the GNN evaluation was run on an NVIDIA Titan RTX GPU), we are nevertheless confident that the GNS evaluation compares favorably even with an optimized and parallelized FEM code.

\section{Conclusions and Future Work}

A tool to accelerate thermal simulations using GNNs has been presented. This could enable fast evaluation of the thermal performance of candidate designs in an SBD process. Our GNS is trained on a large set of FEM simulations of 3D systems. The main contribution of this work is that the 3D systems used for training are much simpler than any realistic electronic system and, hence, the computational cost of generating the training dataset is low. Nevertheless, the learned GNS is still applicable to realistic systems and achieves a remarkably low one-step and accumulated rollout error in a time arguably shorter than any FEM simulation tool. On-going work focuses on reducing the accumulated unrolling error even further and on improving the training systems in order to minimize to associated computational costs.

This work considers heat propagation via conduction only. In a realistic situation, heat rejection in an electronic system will also take place via (forced) convection. Developing a learned GNS that can deal with both heat conduction and convection will be the subject of future work.

\section*{Acknowledgment}

This work has been supported by Silicon Austria  Labs GmbH (SAL), owned by the Republic of Austria, the Styrian Business Promotion Agency (SFG), the federal state of Carinthia, the Upper Austrian Research (UAR), and the Austrian Association for the Electric and Electronics Industry (FEEI).


\bibliographystyle{IEEEtran} 
\bibliography{gnn_biblio}

\end{document}